\documentclass[aps,prl,twocolumn,showpacs,preprintnumbers]{revtex4}
\usepackage{amsmath,amssymb}
\usepackage{bm}
\usepackage{comment}
\renewcommand\d{\partial}

\newcommand\Dlr{\raisebox{0.1em}{$\stackrel{\scriptstyle\leftrightarrow}D$}}

\preprint{EFI-14-4}
\begin{document}

\title{Effective Field Theory of Relativistic
  Quantum Hall Systems} 

\author{Siavash Golkar, Matthew M. Roberts, and
  Dam Thanh Son} \affiliation{Kadanoff Center for Theoretical Physics,
  University of Chicago, Chicago, Ilinois 60637, USA}

\begin{abstract}
Motivated by the observation of the fractional quantum Hall effect in
graphene, we consider the effective field theory of relativistic
quantum Hall states.  We find that, beside the Chern-Simons term, the
effective action also contains a term of topological nature, which
couples the electromagnetic field with a topologically conserved
current of $2+1$ dimensional relativistic fluid.  In contrast to the
Chern-Simons term, the new term involves the spacetime metric in a
nontrivial way.  We extract the predictions of the effective theory
for linear electromagnetic and gravitational responses.  For
fractional quantum Hall states at the zeroth Landau level, additional
holomorphic constraints allow one to express the results in terms of two
dimensionless constants of topological nature.
\end{abstract}

\maketitle

\emph{Introduction.---}Recently both
integer~\cite{Novoselov:2005kj,Zhang:2005zz} and
fractional~\cite{FQHE-graphene1,FQHE-graphene2} quantum Hall effects
have been observed in graphene.  In many respects, graphene behaves
like a relativistic system; in particular the integer quantum Hall
(IQH) plateaux corresponds to the Hall conductivity (in unit of the
quantum $(e^2/h$) $\nu=4(n+\frac12)$, with the factor 4 due to the
spin and valley degeneracies and the offset of $1/2$ due to the
relativistic nature of the low-energy electron spectrum near the Dirac
points.  Fractional quantum Hall (FQH) states in graphene have
also been observed; in particular, many standard fractions in the
range $0<\nu<1$ have been seen.

Motivated by the graphene, in this paper we
study the relativistic version of the quantum Hall effect.  Although
electrons in graphene do not behave in a strictly Lorentz invariant
fashion (the Coulomb interaction is practically instantaneous), one
can still hope to draw physically relevant lessons for graphene when
the physics is insensitive to the velocity of propagation of
the interaction.  We will see that Lorentz invariance imposes
stringent constraints on the low-energy behavior of the system, which
enable one to express a multitude of physical observables through a
small number of parameters.  In particular, we find that there are two
topological parameters that enter the effective field theory
description: the Hall conductivity $\nu$, and a relativistic version
of the shift, denoted here as $\kappa$.  The latter is defined as the
offset between the total charge $Q$ and the magnetic flux in units of
the flux quantum $N_\phi$ when the system is put on a sphere with no
quasiparticles or quasiholes present: $Q=\nu N_\phi+\kappa$.  Together
with a single, nonuniversal function of the magnetic field, denoted as
$f(b)$, all parity-odd transports of the quantum Hall system are
completely determined up to second order in the expansion over
momentum.  In particular, the system possesses a Hall
viscosity~\cite{Avron:1995fg} proportional to $\kappa$.  This
relationship is the relativistic analog of the exact relationship
between the Hall viscosity and the shift in nonrelativistic
systems~\cite{ReadRezayi:2010}.

There is a further simplification for quantum Hall states on the
zeroth Landau level (which corresponds to $-2\le\nu\le2$ in graphene),
in the limit of negligible Landau-level mixing.  The holomorphic
constraint on the low-energy states determines completely the function
$f(b)$ which is otherwise non-universal.  In this case, all parity-odd
transport is
universal to second order in the expansion over momentum.  For
example, the frequency and momentum dependence of the Hall
conductivity is found to be
\begin{equation}\label{sigmaxy}
  \sigma_{xy}(\omega,q) = \frac\nu{2\pi}\Bigl(1+\frac{2\omega^2}{\omega_c^2}
  \Bigr)
    + \frac{\kappa-\nu}{8\pi}(q\ell_B)^2,
\end{equation}
where $\ell_B=\sqrt{\hbar c/eB}$ is the magnetic length and
$\omega_c=v_F/\ell_B$, with $v_F$ being the Fermi velocity.
Equation~(\ref{sigmaxy}) parallels a similar formula in the
nonrelativistic case~\cite{Hoyos:2011ez,Bradlyn:2012ea}.  The
dependence on the shift is exactly the same, after the identification
$\kappa\sim\nu{\cal S}$, but there is a difference in a constant in
front of $(q\ell_B)^2$.  This is the reflection of the fact that the
electromagnetic current is not a simple lowest-Landau level operator.

\emph{Power-counting.}---From now on we set $\hbar=v_F=1$ and absorb
the electron charge $e$ into the magnetic field.  The low-energy
dynamics in the bulk of a gapped quantum Hall system can be described by
a local effective action which is a functional of the external probes.
We will turn on both the electromagnetic and gravitational
perturbations.  We regard the external magnetic field $B$ to be
$O(1)$ and consider perturbations of the external gauge field that are
of the same order as the background: $F_{\mu\nu}=O(1)$.  Denoting the
momentum scale by $p$, one then has $A_\mu=O(p^{-1})$.  The
perturbations of the metric is assume to be of order one:
$g_{\mu\nu}=O(1)$, so, for example, the Riemann tensor is $O(p^2)$.

The term in the effective Lagrangian with the lowest power of $p$ is
the Chern-Simons term,
\begin{equation}
  S_{\rm CS} = \frac\nu{4\pi}\int\!d^3x\sqrt{-g}\,\epsilon^{\mu\nu\lambda}
   A_\mu \d_\nu A_\lambda,
\end{equation}
and is $O(p^{-1})$.  At the next, $O(1)$, order, there is only one
gauge invariant scalar $F_{\mu\nu}F^{\mu\nu}$, so the most general
contribution to the action at this order is
\begin{equation}
  S_\epsilon = -\int\!d^3x\,\sqrt{-g} \, \epsilon(b),
\end{equation}
where $b = ( \frac12 F_{\mu\nu} F^{\mu\nu})^{1/2}$ and $\epsilon$ can
be any function of $b$.  We also introduce a unit timelike vector
$u^\mu$, $u^2=-1$, defined through
\begin{equation}
  u^\mu = \frac1{2b}\epsilon^{\mu\nu\lambda} F_{\nu\lambda},
\end{equation}
which corresponds to the local frame in which the electric field
vanishes.  The Bianchi identity $dF=0$ becomes $\nabla_\mu(bu^\mu)=0$.
The action $S_\epsilon$ depends on the metric, and varying the action
with respect to the metric one finds that the stress-energy tensor is,
to this order $T^{\mu\nu}=(\epsilon+P)u^\mu u^\nu +Pg^{\mu\nu}$, where
$P=b\epsilon'(b)-\epsilon(b)$.  The stress-energy tensor has the form
of that of an ideal fluid.

To construct higher-order terms, it is convenient to use $b$ and
$u^\mu$ instead of $F_{\mu\nu}$.  Both $b$ and $u^\mu$ are $O(1)$ in
our power-counting scheme.  The definition of $u^\mu$ involves $b$ in
the denominator, but this does not create any problem since the
effective theory is supposed to work only at finite magnetic field.
The presence of the unit vector $u^\mu$ reminds us of the
Einstein-aether theory~\cite{Jacobson:2000xp}, but there are
importance difference due to the fact that we are in (2+1) dimensions.
One can write down two obvious terms at order $O(p)$,
\begin{equation}
  f(b)\epsilon^{\mu\nu\lambda} u_\mu\d_\nu u_\lambda, \quad
  f_2(b) u^\mu \d_\mu b .
\end{equation}
However, by introducing a new function $f_3(b)$ so that
$f_2(b)=bf_3'(b)$, the second term can be shown to vanish by
integration by parts.  There exists, however, one more term at order
$O(p)$.  The construction of this term involves a topological current
which we now describe.

\emph{Topological current.---}One notices that the following current
is identically conserved,
\begin{equation}\label{J}
   J^\mu =  \frac1{8\pi} \varepsilon^{\mu\nu\lambda} 
    \varepsilon^{\alpha\beta\gamma} 
    u_\alpha \Bigl(\nabla_\nu u_\beta \nabla_\lambda u_\gamma -
    \frac12 R_{\nu\lambda\beta\gamma} \Bigr).
\end{equation}
The current has a topological interpretation. The total charge calculated on any space-like surface is simply the Euler character of this surface. The only exception is the special case of a Euclidean
space time containing an $S^2$, where the total charge becomes the
Euler character multiplied by the winding number of the $S^2\to S^2$
map from the 2D spatial slice to the space of unit vector
$u^\mu$ (this winding number in Lorentz signature is equal to 1). We
can motivate this in the simple case when $u^\mu\sim (1,\vec{0})$. 
Then the charge density associated with the
current~(\ref{J}) is the scalar curvature of the 2D surface of constant time,
\begin{equation}\label{JR}
  J^0 = \frac1{8\pi} {}^{(2)}\!R,
\end{equation}
so the total charge is proportional to the Euler characteristic $\chi$
of the hypersurface, $Q=\chi/2=1-g$, where $g$ is the genus of the
surface. A more thorough investigation of this new topological current will be presented in \cite{TBA}.

\emph{The second topological term.}---We can now add to the effective
action a term $\kappa\!\int\!d^3x\sqrt{-g}\,A_\mu J^\mu$.  Since
$J^\mu$ is identically conserved, this term is gauge invariant up to a
boundary term, similar to the Chern-Simons term.  But in contrast to
the Chern-Simons term, the new term exists only for background where
the magnetic field does not vanish anywhere.  This restriction is
natural for the effective field theory of a quantum Hall state.
Moreover, the term is $O(p)$ in our power counting
scheme, thus we have to take that into account when working to
this order.  

The term under consideration is the relativistic counterpart of the
mixed Chern-Simons term $A\wedge d\omega$ which appears in the
nonrelativistic case~\cite{Wen:1992ej,Son:2013rqa}.  In the (2+1)D
relativistic theory where the spin connection is non-Abelian such a
term cannot be directly written down.

Thus, the final action, including all terms to order $O(p)$ is
\begin{multline}\label{L-eff}
  \mathcal L = \frac\nu{4\pi} \varepsilon^{\mu\nu\lambda}
  A_\mu\d_\nu A_\lambda - \epsilon(b) 
  + f(b) \varepsilon^{\mu\nu\lambda}
  u_\mu\d_\nu u_\lambda\\
  + \frac\kappa{8\pi}
  \varepsilon^{\mu\nu\lambda}\varepsilon^{\alpha\beta\gamma} A_\mu
  u_\alpha\Bigl(\nabla_\nu u_\beta \nabla_\lambda u_\gamma - \frac12
  R_{\nu\lambda\beta\gamma} \Bigr). 
\end{multline}

It is interesting to note that when $\epsilon(b)\sim b^{3/2}$ and
$f(b)\sim b$, the action is fully Weyl invariant.  This would be the
case if the microscopic theory underlying the quantum Hall state is a conformal field theory.

 \emph{Relativistic shift.}---The
coefficient $\kappa$ is related to a relativistic version of the
shift~\cite{Wen:1992ej}.  The charge density is the
variation of the action with respect to $A_0$. One can see that the
total charge on a closed surface comes only from the Chern-Simons and
the $\kappa$ term in the Lagrangian~\eqref{L-eff},
\begin{equation}
  Q = \int\!d^2x\Bigl(\frac\nu{2\pi} F_{12} + \frac\kappa{8\pi} J^0\Bigr).
\end{equation}
By using~(\ref{JR}) this can be written as $Q=\nu
N_\phi+\kappa\chi/2$, where $N_\phi$ is the total number of magnetic
flux quanta threading the manifold and $\chi$ is the Euler character
of the manifold.  This relationship is a relativistic version of the
the shift, which is normally defined as $\mathcal{S}$ in the equation
$Q=\nu(N_\phi+\mathcal{S})$ on a sphere \cite{Wen:1992ej}.  We have defined $\kappa$ to remain
finite at $\nu=0$.

For an integer quantum Hall states with $\nu=N_f(n+\frac12)$, where
$N_f$ is the total number of ``flavor'' degeneracy of the Landau
levels (in graphene $N_f=4$) the total charge can be found by summing
up all charges of the filled Landau levels on a sphere.  We find
$\kappa=N_f n(n+1)$.  Note that $\kappa=0$ for the $\nu=\pm2$ states
in graphene.

For fractional quantum Hall states in graphene the value of $\kappa$
is related to the shift $\mathcal{S}$ of the corresponding state in
the usual nonrelativistic theory.  For illustration let us consider a
state with $0<\nu<1$ in graphene with complete SU(4) breaking.  Among
the four zeroth Landau levels, two are completely filled, and a third
zeroth Landau level is incompletely filled, and the fourth is
completely empty.  The lowest Landau level of the Dirac fermion on a
sphere with $N_\phi$ magnetic flux quanta threading it, is identical to
the Landau levels of a nonrelativistic fermion on sphere with
$N_\phi'=N_\phi-1$ magnetic flux quanta (in particular it has
degeneracy $N_\phi=N_\phi'+1$).  For the latter, the formula
$Q=\nu(N_\phi'+\mathcal S_\text{NR})$ is taken as the definition of
$\mathcal S_\text{NR}$,
which implies $\kappa=\nu(\mathcal{S}_\text{NR}-1)$.  For example, the
$\nu=\frac13$ state has $\kappa=\frac23$. 

\emph{Discrete symmetries.}---Let us assume the the microscopic theory
respects $C$, $P$, and $T$, and discuss if the effective theory breaks
these symmetries.  Recall that~\cite{Deser:1981wh} under $C$,
$A_\mu\to-A_\mu$; under $P$, $x^1\to-x^1$, $A_0\to A_0$, $A_1\to-A_1$,
and $A_2\to A_2$; and under $T$ $t\to-t$, $A_0\to A_0$ and
$A_i\to-A_i$.  All these symmetries are broken by the background
magnetic field, and $C$ is further broken if there is a nonzero
chemical potential.  But one combination, $PT$, leaves both the
magnetic field and the chemical potential invariant.  It is easy to
see that all terms we have considered are invariant with respect to
$PT$.  When the chemical potential is zero, however, we can classify
our terms also with respect to $CP$ and $CT$.  The latter is the
particle-hole symmetry of the lowest Landau level. Under these
combinations $\nu$, $\kappa$, and $f(b)$ all change signs.  Therefore
the presence of a nonzero $\nu$, $\kappa$, or $f(b)$ at zero chemical
potential signals a spontaneous breaking of $CP$ and $CT$ symmetries.
The $\nu=0$ IQH state of graphene has $\kappa=0$, consistent with
unbroken $CP$ and $CT$.  On the other hand, it is easy to construct a
multiflavor Moore-Read state~\cite{Moore:1991ks} at $\nu=0$ which
breaks these symmetries.

\emph{Momentum density.}---As the first application of the effective field theory, we compute the momentum density
$T^{0i}$ in the background of inhomogeneous magnetic field $B=b(x,y)$.  
To that end, we 
turn on a perturbation in the $g_{0i}$ component of the metric tensor
and read out the momentum density from the action: $\delta S =
\int\!d^3x T^{0i}g_{0i}$.  We find
\begin{equation}
	\label{eq:mom_density}
  T^{0i} = - \epsilon^{ij}\d_j \left(\frac\kappa{8\pi}b+f(b)\right) .
\end{equation}


\emph{FQH states on the zeroth Landau level.}---We now show that, for
the FQH states on the zeroth Landau level with negligible mixing with
other Landau levels, the function $f(b)$ is completely determined by
the topological coefficients $\nu$ and $\kappa$.  This comes from a
holomorphic constraint relating the momentum density and the particle
density.

For concreteness, we choose the following representation for the
$2\times2$ Dirac matrices
\begin{equation}
  \gamma^0 = \sigma_3, \quad \gamma^1 = i\sigma_2, \quad
  \gamma^2 = -i\sigma_1,
\end{equation}
for which the free Hamiltonian has the form
\begin{equation}
  H = -i \gamma^0\gamma^i D_i - A_0 = -2i\begin{pmatrix}
      0 & D \\ \bar D & 0 \end{pmatrix} - A_0.
\end{equation}
Here $D\equiv D_z$, $\bar D\equiv D_{\bar z}$, and we use complex coordinates:
$z = x+iy$, $\bar z = x-iy$. 
The $n=0$ Landau level are $\psi = (\varphi,0)^T$, where $\varphi$ satisfies
the holomorphic constraint $\bar D \varphi =\bar D \varphi^* = 0$.  
Now let us look at the stress-energy tensor,
\begin{equation}\label{Tmunu}
  T^{\mu\nu} = -\frac i4 \bar\psi \gamma^{(\mu} \Dlr{}^{\nu)} \psi,
\end{equation}
assuming a static, but spatially inhomogeneous, magnetic field, and no
electric field.  For the $0i$ components we can ignore time
derivatives as $A_0=0$ and the lowest Landau level has zero energy.
We see that
\begin{equation}
  T^{0i} = -\frac i4 \varphi^* \Dlr{}^i \varphi,
\end{equation}
which, by using the holomorphic constraints, can be transformed into
\begin{equation}
  T^{0i} = -\frac14 \epsilon^{ij}\d_j n,
\end{equation}
where $n=\varphi^*\varphi$ is the particle number density on the
lowest Landau level.  Comparing that to \eqref{eq:mom_density}, we find
that, for FQH states in the zeroth Landau level,
\begin{equation}
  f(b) = \frac1{8\pi}(\nu-\kappa) b.
\end{equation}
The calculation above neglects possible mixing between Landau levels,
as well as the possible corrections to the stress-energy
tensor~(\ref{Tmunu}) due to interactions.  Both effects are small when
the interaction energy scale is much smaller than the distance between
Landau levels $\sqrt B$.

\emph{Response functions.}---We now compute different response
functions of the relativistic quantum Hall states to external fields.

First we compute the Hall viscosity.  The Hall viscosity is defined
through response to uniform shear metric perturbations.  For
simplicity we turn on only spatially homogeneous perturbations of the
spatial components of the metric, $g_{ij}=\delta_{ij}+h_{ij}(t)$.  The
relevant term in the action is
\begin{equation}
 \kappa \int \!d^3x \,\sqrt{-g} A_\mu J^\mu = 
  	-\frac{\kappa B}{32 \pi} 
  	\int \!d^3x \, \epsilon^{jk} h_{ij} \d_t h_{ik},
\end{equation}
where we have performed integration by parts.
We find
\begin{equation}
  \eta_H = \frac{\kappa B}{8 \pi}\,.
\end{equation}
The relationship between the Hall viscosity and $\kappa$ is identical to the nonrelativistic result $\eta_H = n \mathcal{S}/4$ \cite{ReadRezayi:2010} with the substitution $\mathcal{S} \rightarrow \mathcal{S}_{\text{NR}}-1$ for grapheme states with $0<\nu<1$.
Note that the Hall viscosity depends only on the topological number
$\kappa$.  It is natural since $\eta_H$ can be determined by adiabatic
transport and hence should not depend on non-universal functions like
$f(b)$.

Next we look at the components of the stress tensor when one turns
on a static, spatially inhomogeneous, electric field.  The result is
\begin{equation}
\begin{split}
  T_{ij} = P\delta_{ij} &+ \frac{\kappa}{8\pi} (\d_i E_j + \d_j E_i) \\
  	&- \left(\frac{\kappa}{4\pi}+f'(B)\right) \delta_{ij} 
  \bm{\nabla}\cdot {\bf E},
\end{split}
\end{equation}
which can be written in terms of the drift velocity
$v^i=\epsilon^{ij}E_j/B$ and the shear rate
$V_{ij}=\frac12(\d_iv_j+\d_jv_i -\delta_{ij}\d\cdot v)$,
\begin{equation}
\begin{split}
  T_{ij} = P\delta_{ij} &- \eta_H (\epsilon_{ik}V_{kj}+\epsilon_{jk}V_{ki})\\
   & + \delta_{ij} (\eta_H+Bf'(B)) \bm{\nabla}\times {\bf v}.
\end{split}
\end{equation}
The traceless part of the stress tensor reflects the nonzero Hall
viscosity of the quantum Hall
fluid~\cite{Avron:1995fg,Avron:1997,Read:2008rn}.

The two point functions of currents give us the response to external
electromagnetic field, derived from the quadratic part of
the effective action in flat space,
\begin{equation}
  \mathcal L = -\left( \frac\kappa{8\pi B} + \frac{f(B)}{B^2}\right) 
    \epsilon^{ij}E_i \d_t E_j - \frac{f'(B)}B E_i \d_i B.
\end{equation}
In particular, we find the correction to Hall conductivity
$\sigma_{xy}$ (for longitudinal electric fields) at nonzero
frequencies and wavenumbers,
\begin{equation}
  \sigma_{xy}(\omega,q) = \frac\nu{2\pi} +
   \left( \frac\kappa{4\pi} + \frac{2f(B)}B \right) \frac{\omega^2}B
   - \frac{f'(B)}B \frac{q^2}B\,.
\end{equation}
At lowest Landau level, the formula becomes
\begin{equation}\label{Kohn}
  \sigma_{xy}(\omega,q) = \frac\nu{2\pi} + \frac\nu\pi \frac{\omega^2}B
    + \frac{\kappa-\nu}{8\pi}\frac{q^2}B\,.
\end{equation}

In Galilean invariant systems, the frequency dependence of the
conductivity matrix is completely determined, at $q=0$, by Kohn's
theorem~\cite{Kohn:1961zz}.  In relativistic systems Kohn's theorem no
longer applies.  Nevertheless, Eq.~\eqref{Kohn} implies that, in
Lorentz invariant systems, the $\omega^2$ correction is completely
fixed by the filling fraction.

We now discuss the relevance of our formulas for graphene, where
the Lorentz invariance of the low-energy free theory is broken by the
Coulomb interaction~\cite{Gonzalez:1993uz}.  In the limit of weak
Coulomb interaction our formula should work, to leading order of the
interaction strength, for integer quantum Hall states.  For fractional
quantum Hall states in the zeroth Landau level, if one is interested
in the response for $\omega\ll q$, the effect of retardation of the
interaction should be small.  In this case, one can replace the
instantaneous Coulomb interaction by a Lorentz invariant interaction
without changing the response functions.  Thus the $q^2$ correction to
$\sigma_{xy}$ is reliable for graphene FQH states at the zeroth Landau
level.

\emph{Summary.}---We constructed an effective field theory description
for relativistic quantum Hall liquids.  The theory contains one
additional topological coefficient besides the Hall conductivity.  For
states at the zeroth Landau level, there is an additional holomorphic
constraint which completely determines the effective Lagrangian in
terms of the two topological numbers.  Our formalism elucidates the
intricate relationship between topology and geometry in the problem.
In particular, there is an important term in the Lagrangian with a
topologically determined coefficient ($\kappa$), but the term itself
depends nontrivially on the metric of space.

It would be interesting to extend the theory to higher orders in
momentum expansion and compare the predictions for electromagnetic and
gravitational responses to the nonrelativistic
case~\cite{Can:2014ota,Abanov:2014ula}.

We note that on a manifold with a boundary, conservation  of the topological current requires the addition of a boundary action. However, this is only possible if the field $u_\mu$ is parallel to
the boundary. If this is not the case, the gauge non-invariance of the
action should be absorbed by the anomaly of the boundary theory.  The
implication of this should be
further investigated.

Finally, our effective description may serve as a benchmark for
holographic models of quantum Hall effect, many of which have
underlying Lorentz invariance~(see, e.g., Ref.~\cite{Bergman:2010gm}).

The authors thank Nicholas Read and Paul Wiegmann for discussions.
This work is supported, in part, by DOE grant DE-FG02-13ER41958.
S.G. is supported in part by NSF MRSEC grant DMR-0820054.  D.T.S. is
supported in part by a Simons Investigator grant from the Simons
Foundation.

\end{document}